\documentclass[aps,prl,preprint,superscriptaddress,showpacs]{revtex4}

\usepackage{amssymb}
\usepackage{epsfig}
\usepackage{amsmath}
\usepackage{times}

\begin{document}

\title{Visualizing and exploring modular networks based on a probabilistic model}



\author{Xiaofeng Gong}
\affiliation{Temasek Laboratories, National University of Singapore, Singapore 117508}
\affiliation{Beijing-Hong Kong-Singapore
Joint Center of Nonlinear and Complex systems (Singapore), National University of Singapore, 
Singapore 117508}
\author{C.-H. Lai}
\affiliation{Department of Physics, National University of Singapore, Singapore 117542}
\affiliation{Beijing-Hong Kong-Singapore
Joint Center of Nonlinear and Complex systems (Singapore), National University of Singapore,
Singapore 117508}

\begin{abstract}
	We propose a method to investigate modular structure in networks based on fitted probabilistic model, where the connection probability between nodes is related to a set of introduced local attributes. The attributes, as parameters of the empirical model, can be estimated by maximizing the likelihood function of the observed network. We demonstrate that the distribution of attributes provides an informative visulization of modular networks on low-dimensional space, and suggest the attribute space can be served as a better platform for further network analysis. 

\end{abstract}

\maketitle

Networks are widely used to model complex systems with many interactive units\cite{Barabasi2002,Holme2004}. Usually, each node in such a network represents a distinct individual, and a link is established based on certain measurement of interaction between particular pair of nodes. On the microscopic level, the underlying system is fully described by state of each unit defined by several local properties, which also determines the interactions among them through complicated coupling. Thus, the network would be completely determined if local properties of nodes and interactive functions were known. In many applications however, the representative network is the only available data. It is the purpose of the researchers to infer these crutial information on the underlying system through network analysis. For instance, if the interaction between two units mainly depends on their local properties, then it is reasonable to expect that the units which have similar connection patterns in network representation, will share some common features in their local properties, and therefore may have similar functioning in the underlying system. 
Although very friutful, the applicability of this type of analysis is limited by the gap existing between the network-level description and the underlying system. This is due to the fact that in most situations, the representitive network only models relationships or interactions among units, but not the associated properties or states of units which determine interactions. So for example, a minor change on the observed network may not be caused necessarily by small perturbations on local properties of units in the underlying system, since interactions may depend on local states in a highly nonlinear way. Similarly, the evolution of the unerlying system due to continuous change of local states may result in abrupt changes on the network structure, such as group merge or split. 

To better deal with these problems, a necessary step is to bridge these two deffierent description levels. However it is generally impossible to completely reconstruct the intrinsic properties of nodes based solely on the strucure of a given network, since the mechanism determining a link may be very complicated.
In this paper, we develope a method to describe the system in a middle level between the representitive network and the microscopic description through some simplifications. 
We regard the representitive network as one particular realization of an emprical probabilistic model. In this model, each pair of nodes has certain probability to be connected, and the connection probabilities depends on the distance between local attributes of the corresponding nodes through some function. The attributes, act as model parameters, can be estimated according to some statistical criterion to best interpret the observed connections. Since each data point in the space of attributes is associated with distinct node, their configuration provides an alternative representation of the observed network.
 
The advantages of introducing this new representation are twofolds. 
On the one hand, the configuration of the attributes provides an informative projection of the network on a low-dimensional space. The established attribute space can be taken as new platform for further network analysis, where the attribute vectors which have no difference with conventional data source, allow many well developed clustering techniques to be applied directly. On the other hand, the introduced local attributes are closely related to the unknown properties of the unerlying system by the common observed network. It may be easier to model evolution of the underlying system as changes of attributes, and study the final influences on network structure through the empiral probability model. 

In this paper, we mainly focus on studying modular strucutres in networks. After describing technique details of modeling approach, we apply the proposed method to some artificial and real-world networks to demonstrate its usefulness on network visualzing and structure analysis. We also discuss a possible way to extend the proposed technique to deal with more complicated multi-layered modular network.

Assume there is an imaginary probabilistic system with $N$ nodes, which is characterized by a connection probability matrix $C(N \times N)$, where $c_{ij}$ represents the probability of connection between node $i$ and $j$. The given network described by adjancency matrix $A(N \times N)$ is regarded as a realization of this imaginary system. In particular, $A_{ij}$ is treated as a random variable and its observed value is determined by a Bernoulli trial according to the probability $c_{ij}$. For each node in the imaginary system, there associates a set of local attributes denoted by $w_k(1 \times m), k=1,2,\cdots,N$, where the dimension $m \ll N$. The connection probability $c_{ij}$ is then assumed to depend only on the local attributes $w_i$ and $w_j$, and can be written as $c_{ij}=f(w_i,w_j)$. Obviously, this is a great simplication, and should be regarded as first order approximation. 
The function $f$ tying connection probability $c_{ij}$ and local attributes $w_i$ and $w_j$ should be chosen depending on the problem at hand. In our study, we are mainly interested in undirectional unweighted modular networks. Considering the symmetric constraint, we choose a function $f$ which depends only on the distance $\Delta w = \parallel w_i -w_j \parallel$. $f(\Delta w)$ will be closed to $1$ when $\Delta w$ approaching to zeros, and decays rapidly with increasing $\Delta w$ . A natrual choise of $f$ is in Guassian form: $f(w_i,w_j)=e^{-(w_i-w_j)^T\gamma_{ij}(w_i-w_j)}$, where $\gamma_{ij}$, defined as
\[
 \gamma_{ij} = \left\{ \begin{array}{cl}
					1  & \mbox {if degree distribution is homogeneous} \\
					\frac{1}{d_id_j} & \mbox  {if degree distribution is inhomogeneous}
					\end{array},
			\right. 			 
\]
is used as a rescaling factor to compensate the distortion when projecting the high dimensional network into a low-dimensional Euclidian space due to the inhomogeneity of the degree distribution of the network. 
The connection probability thus becomes 
$ c_{ij} = e^{-(w_i-w_j)^T\gamma_{ij}(w_i-w_j)}$.
Therefore, we have 
\begin{eqnarray}
	 p_{ij}      & = & P(A_{ij}=1)  =  e^{-(w_i-w_j)^T\gamma_{ij}(w_i-w_j)}  \\
	\bar{p}_{ij} & = & P(A_{ij}=0)  =  1-P(A_{ij}=1). \nonumber
\end{eqnarray} 
Under this framework, the logarithmic likelihood function of observing particular network $A$ is 
\begin{equation}
	L=\ln Pr(A|w) = \sum_{i,j>i}A_{ij}\ln p_{ij} + \sum_{i,j>i}(1-A_{ij})\ln(1-p_{ij}),
\end{equation}
and
\begin{equation}
	\frac{\partial L}{\partial w_k} \propto -\sum^N_{j=1}\frac{A_{jk}-p_{jk}}{p_{jk}(1-p_{jk})}
	\frac{\partial p_{jk}}{\partial w_k} = -\sum^N_{j=1}\frac{A_{jk}-p_{jk}}{p_{jk}(1-p_{jk})}
	\gamma_{jk}(w_k-w_j).
\end{equation}
Naturally, attributes $w_k$ which maximize above likelihood function are desired.
These maximum likelihood estimates can be obtained by minimizing $-L$ following steps below: 
\begin{enumerate}
	\item starting from arbitary initials $w_0$,
	\item caculate the derivatives $\partial L / \partial w$, 
	\item search in this direction to get $\eta$, so that $w_1=w_0-\eta\frac{\partial L}{\partial w}$ gives smallest $-L$,
	\item set $w_0=w_1$, and go back to step 2.
	\item quit if stopping conditions are met.
\end{enumerate}
It should be noded that the explicitly casted functional relationship of attributes and connection probabilities may not be the true undelrying mechanism. This approach nevertheless, can be applied as a useful technique to explore the desired structure in the network, providing the fitted model is good.

\begin{figure}[htbp]
\begin{center}
\begin{tabular}{cc}
\epsfxsize 7cm \epsfbox{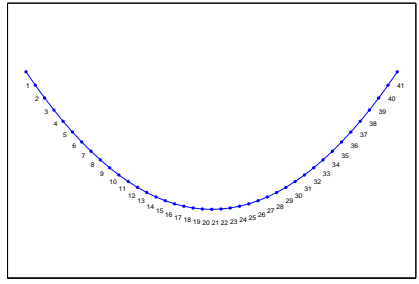} & \epsfxsize 7cm \epsfbox{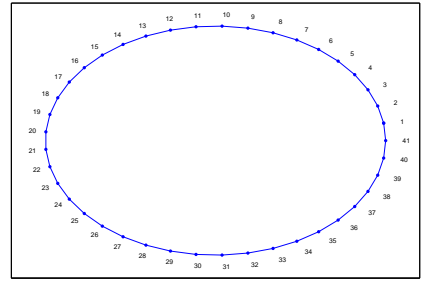} \\
(a) & (b) \\
\epsfxsize 7cm \epsfbox{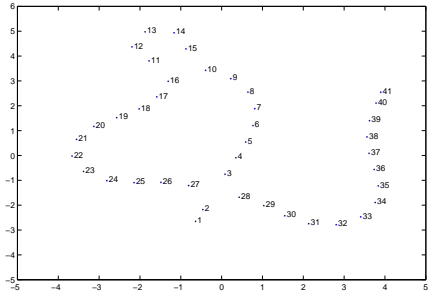} & \epsfxsize 7cm \epsfbox{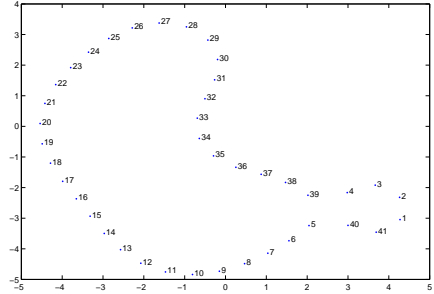}\\
 (c) & (d)
\end{tabular}
\end{center}
\caption{(a) and (b) describe the line and circle networks. (c) and (d) show one typical configuration of attributes corresponding to network in (a) and (b)}
\label{fig:fig1}
\end{figure}

To demonstrate how the distribution of attributes reflects the network topological structure, we first consider two simple networks: a line network and a circle network as shown in figure 1(a) and 1(b). In both networks, one prominent topological feature is the order of the nodes. The essential difference between these two networks lies on the different connection patterns of their ending nodes. Starting from independently random initial distribution ($2D$ Gaussian distribution centered at origin and variance is $4$), the attributes are finally settled down to certain specific configuration after optimization procedure described above. One typical results are shown in figure 1(c) and 1(d) which clearly mirror the  essential structures of both networks. The corresponding nodes are ordered accordingly, and the ending nodes are correctly arranged to capture the different topological features.  The apparent kink(s) observed are caused by the suboptimal (local optimal) nature of the solution generated by gradient based optimization procedure we adopted. In fact, better results (no obvious kinks) can be found seldomly if different initial conditions are tried but not guaranteed. The important point is that global optimal solution may not be necessary to capture the essential topological features. Moreover, if the dimension of the attributes is increased, e.g. form $2D$ to $3D$, the kinks will disappear. However, by this way the model becomes more complex, and its visualization capability is decreased.

Modular netowrks are prevalent in various fields. In some applications, the clustering structure is the only one we are interested in. The proposed probabilistic modeling approach can be applied to modular networks successfully. 
Date points on attribute space generated by the method are usually grouped corresponding to the intrinsic clustering structure, and provides a good visual representation of the investigated network. As it said that a picture is more than thousands of words, in many situations, the picture in attribute space makes it relative easy to get a good guess about the number of clusters and the corresponding partitions even without further analysis.
\begin{figure}[htbp]
\begin{center}
\begin{tabular}{cc}
\epsfxsize 5cm \epsfbox{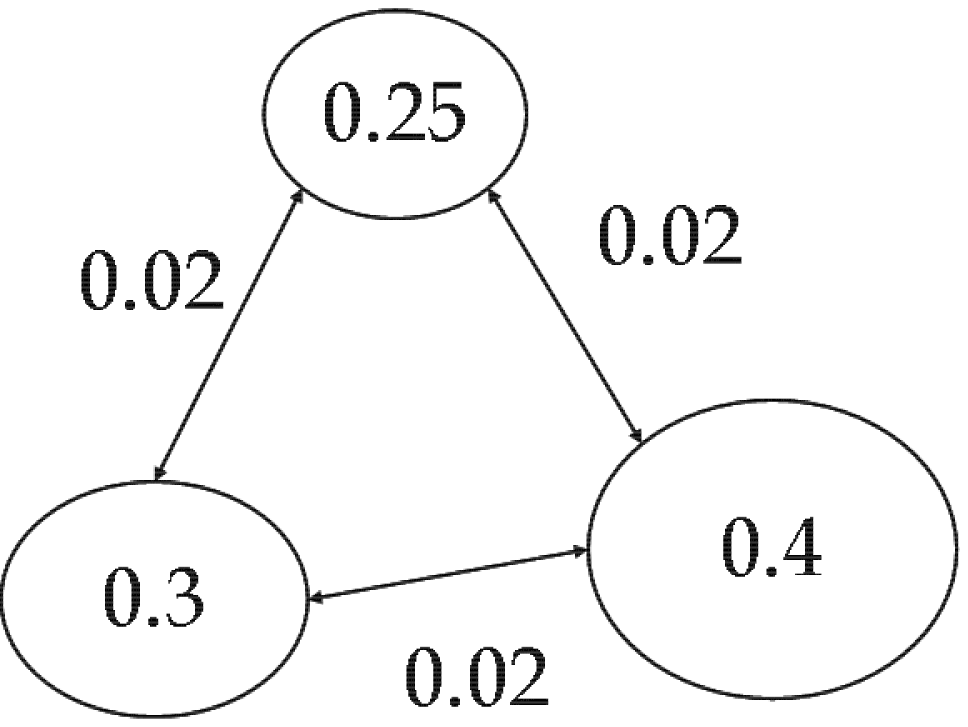} & \epsfxsize 6cm \epsfbox{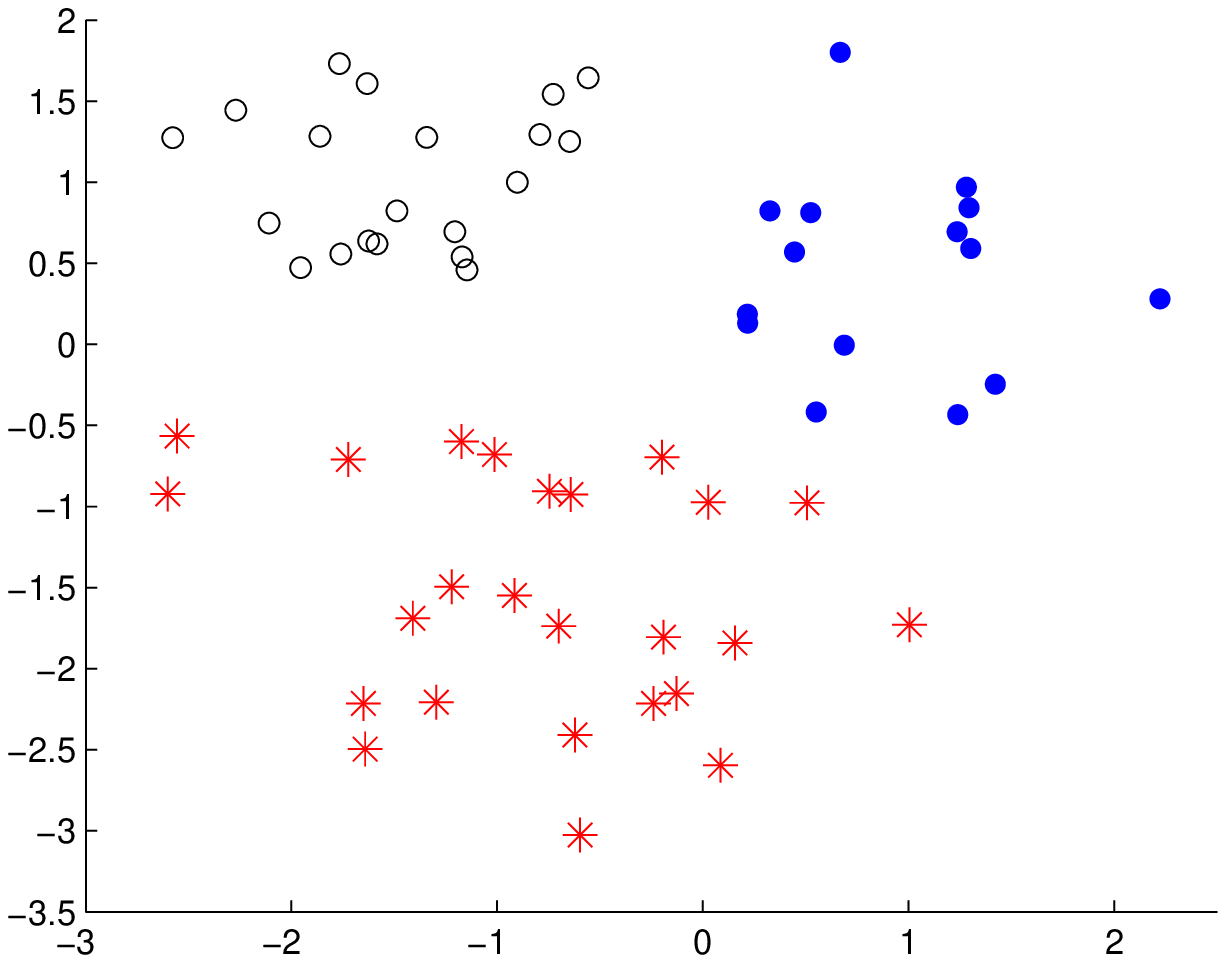}\\
(a)  & (b) \\
\end{tabular}
\end{center}
\caption{(a)Diagram to show the structure of the modular network. The numbers in each circle indicate connection probabilities of nodes in same cluster. The numbers between circles indicate connection probabilities of nodes from different clusters. Three clusters are (nodes $1$ to $25$), (nodes $26$ to $40$) and (nodes $41$ to $60$).  
(b)Attributes configuration on attribute space. Different symbols indicate different clusters. }
\label{fig:fig2}
\end{figure}
\begin{figure}[tbp]
   	\begin{center}
	\epsfig{figure=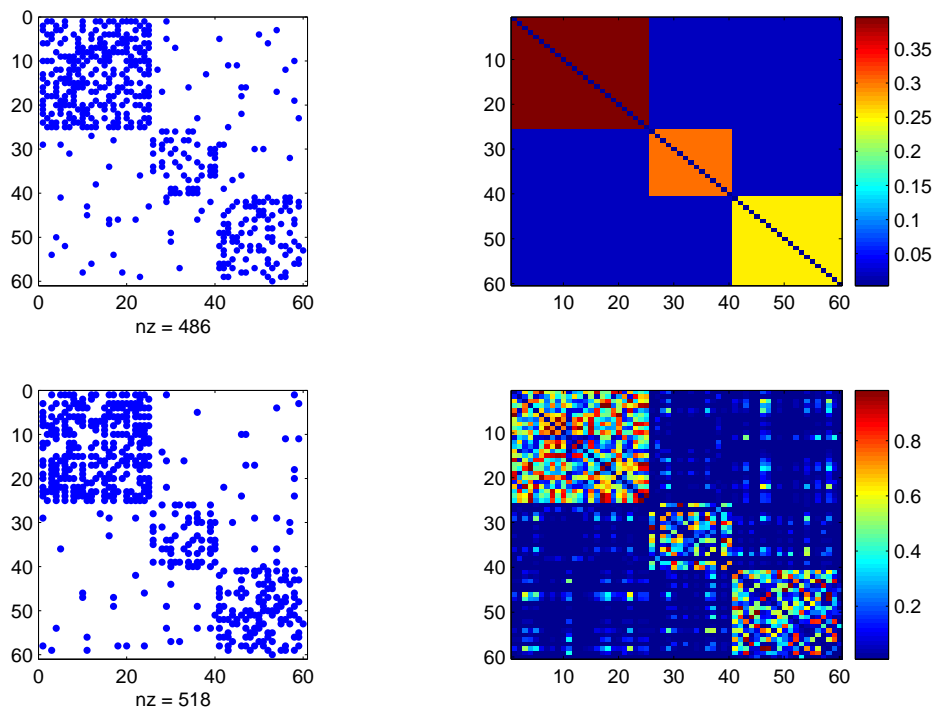,width=1\linewidth}
	\end{center}
     \caption{Comparison of fitted model and true underlying model. Upper part shows the adjacency matrix of observed network (left) and the true underlying probabilistic model which is assumed to be unknown (right). Lower part shows the adjacency matrix of one surrogate network generated by the fitted model (left) and the fitted probabilistic model itself (right).} 
	\label{fig:fig3}
\end{figure}
Figure 2 shows typical simulation results when the method is applied to an artificial clustered networks, which generated from a probabilistic model as described in figure 2(a). The final attributes configuration is shown in figure 2(b), where $3$ clusters are disclosed clearly. The details of fitted model is shown in figure 3 to compare with the true underlying model. The similarity is striking. More careful investigation reveals that the fitting errors are mainly within the clusters. It is more likely that the connection probabilities determined by the fitted model deviate from the true value if the corresponding nodes are in the same cluster than they are in different clusters. Regarding structure analysis, this is actually more desired one comparing to homogeneous fitting errors because connections between nodes in different clusters usually convey much more information on structure than those in the same cluster. 
The main reason for the unsymmetric fitting errors is that the Gaussian function $f$ adopted in our study is not matched with the uniform one in the underlying model. This observation may imply good generalization of proposed modeling approach since exact interactive function is not necessary to get useful results.

The attribute space provides an alternative starting point to make clustering analysis of networks. Since the attribute vectors have no difference with conventional data, there are a lot of well developed clustering techniques can be applied. Furthermore, the informative configuration of attribute vectors not only hint the cluster number, but also provide a good initial partition. With these two key parameters at hand, most clustering algorithms will converge quickly. We apply conventional K-means algorithm \cite{Fukunaga} to above network with initial partitions suggested by the attributes distribution. The algorithm always generate correct partition, and converge quickly after $1$ to $4$ iteratives. 

For real world data, the network structure is more complicated. We apply the proposed approach to several extensively studied real networks, including the karate club network\cite{Zachary1977}, American college football teams network\cite{Girvan2002} and dolphins social network \cite{Lusseau2003}. According to their degree distributions, the rescaling factor $\gamma_{ij}$ is set to be $1$ for football team network, and $1/d_id_j$ for other two networks. The initial attributes distributions are generated by $2$D Gaussian distribution centered at origin and the variance is taken to be $4$. The results for Karate club network are shown in figure 4. It can be seen clearly from figure 4(b) that the distribution of the attributes well reflect the modular structure (shown in figure 4(a)). More details of the fitted model are shown in figure 4(c). One may find that the surrogate network generated by the fitted model is very closed to the observed one.

\begin{figure}[htbp]
   	\begin{center}
	\begin{tabular}{cc}
	\epsfxsize 7cm \epsfbox{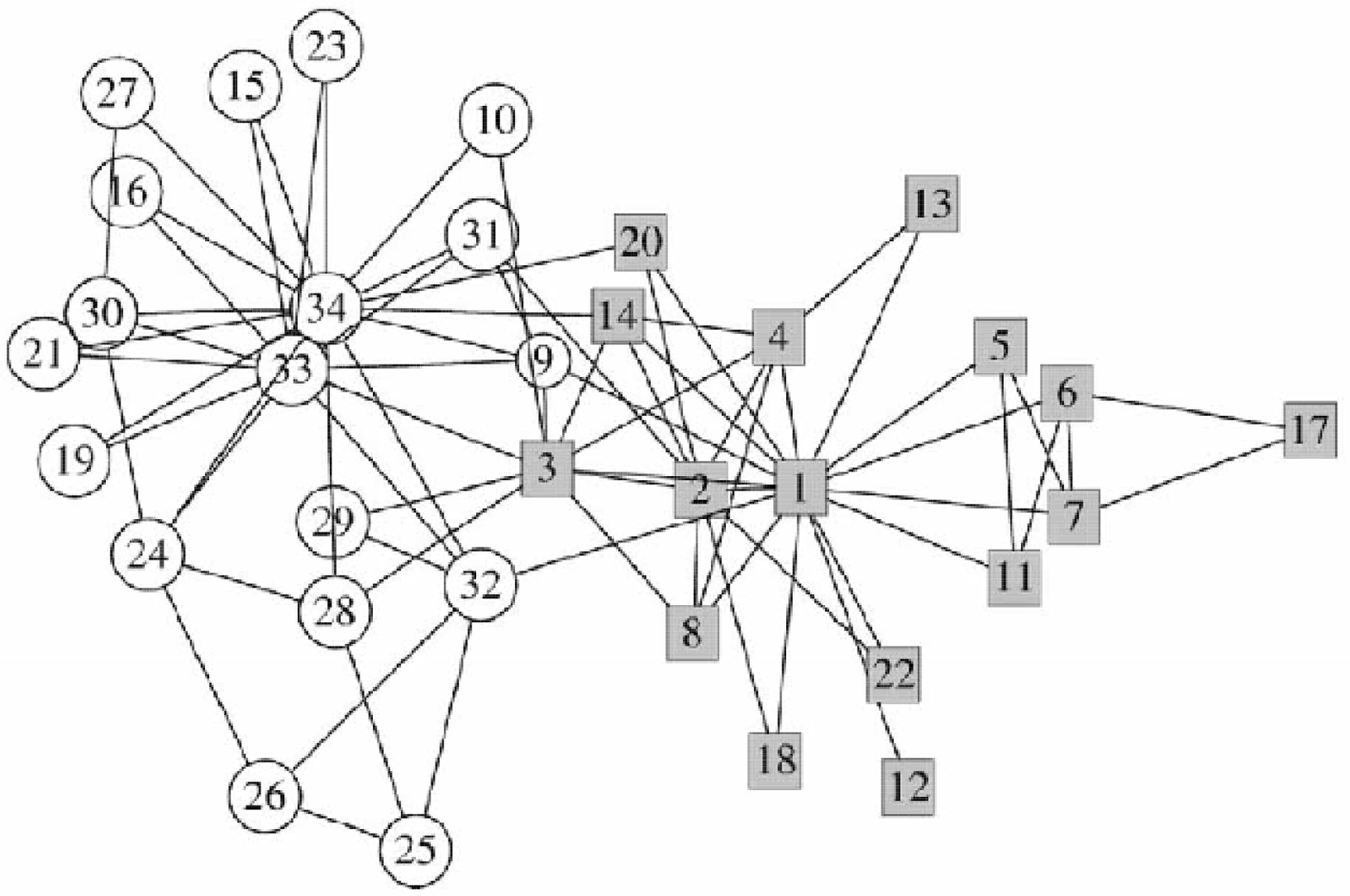} &	\epsfxsize 7cm \epsfbox{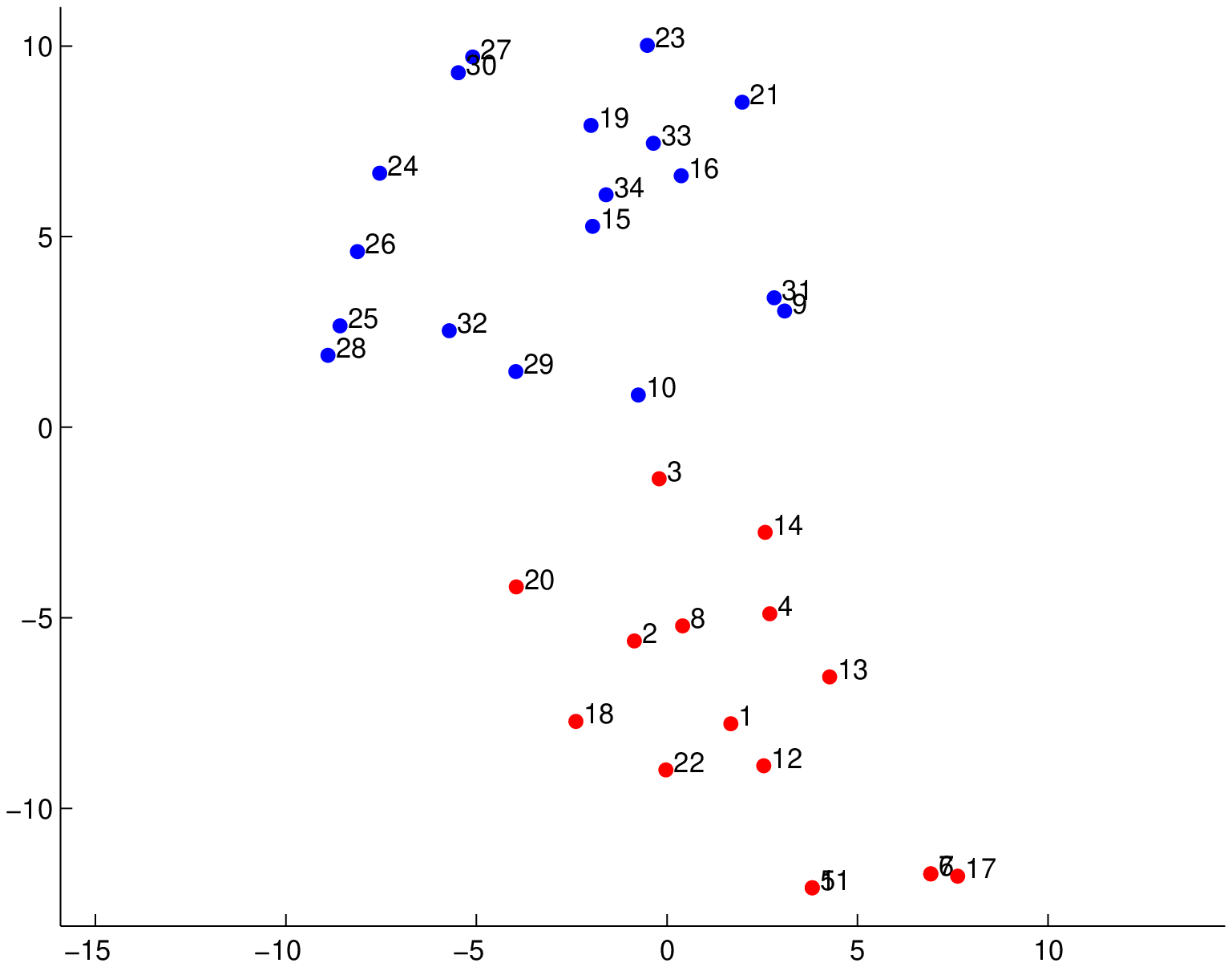}  \\
	(a)  & (b) 
	\end{tabular}
	\begin{tabular}{c}
      \epsfxsize 12cm \epsfbox{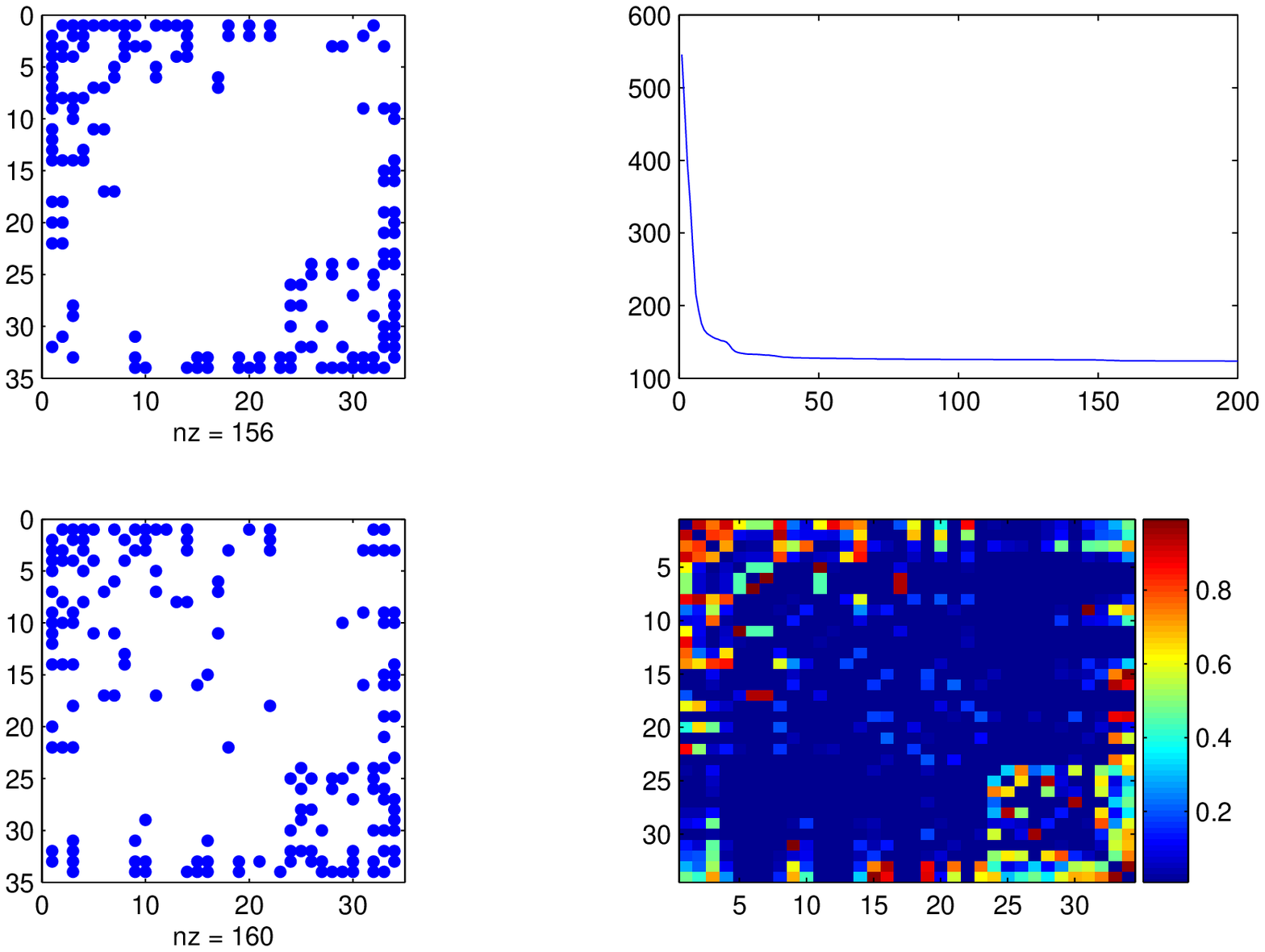}\\
	(c)
	\end{tabular}
	\caption{(a) Karate club network. (b) Distribution of attributes. 
		(c) Fitted probabilistic model of Karate clube network. Upper left: adjacency matrix of Karate club network. Upper right: decay of fitting error when training. Lower left: adjacency matrix of a surrogate network generated by fitted model. Lower right: the fitted probabilistic model.}

	\end{center}
	\label{fig:fig4}
\end{figure}

\begin{figure}[htbp]
   	\begin{center}
	\begin{tabular}{c}
	\epsfxsize 10cm \epsfbox{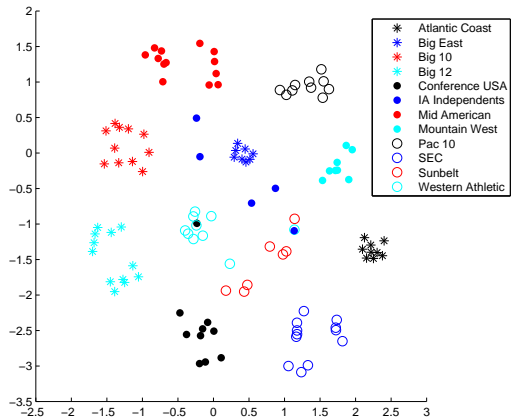}  \\
	(a) \\
      \epsfxsize 10cm \epsfbox{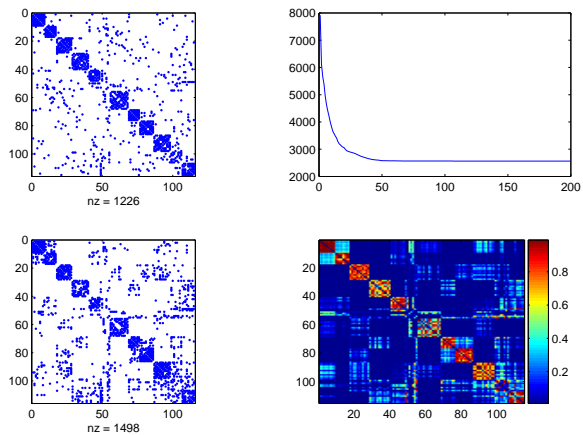}\\
	(b) 
	\end{tabular}
	\end{center}
	\caption{	(a) (colored) Distribution of attributes of American college football teams network. 
			(b) Fitted probabilistic model of American college football teams network. Upper left: adjacency matrix of football teams network. Upper right: decay of fitting error when training. Lower left: adjacency matrix of a surrogate network generated by fitted model. Lower right: the fitted probabilistic model.}
	\label{fig:fig5}
\end{figure}

\begin{figure}[htbp]
   	\begin{center}
	\begin{tabular}{cc}
	\epsfxsize 7cm \epsfbox{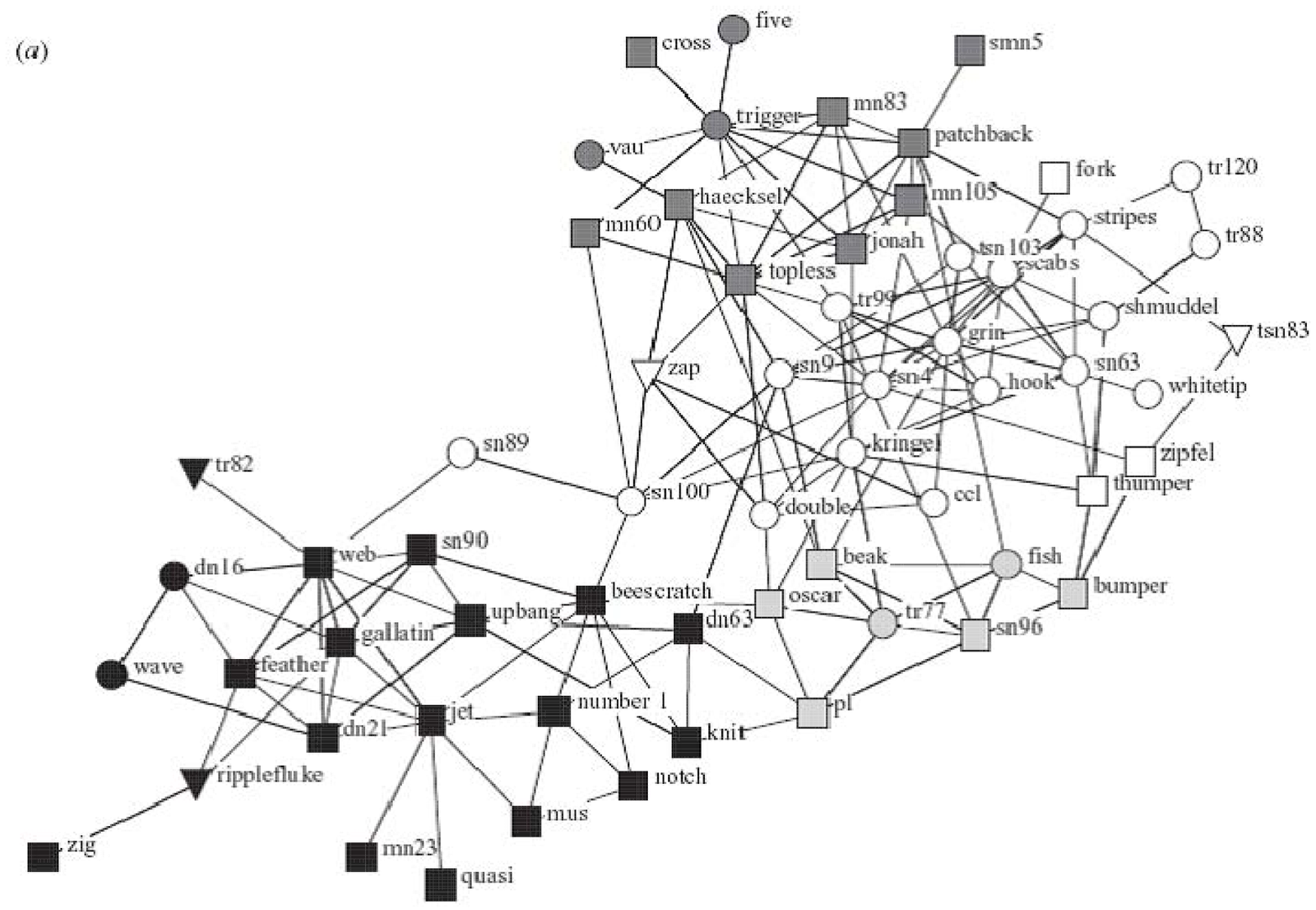} & \epsfxsize 9cm \epsfbox{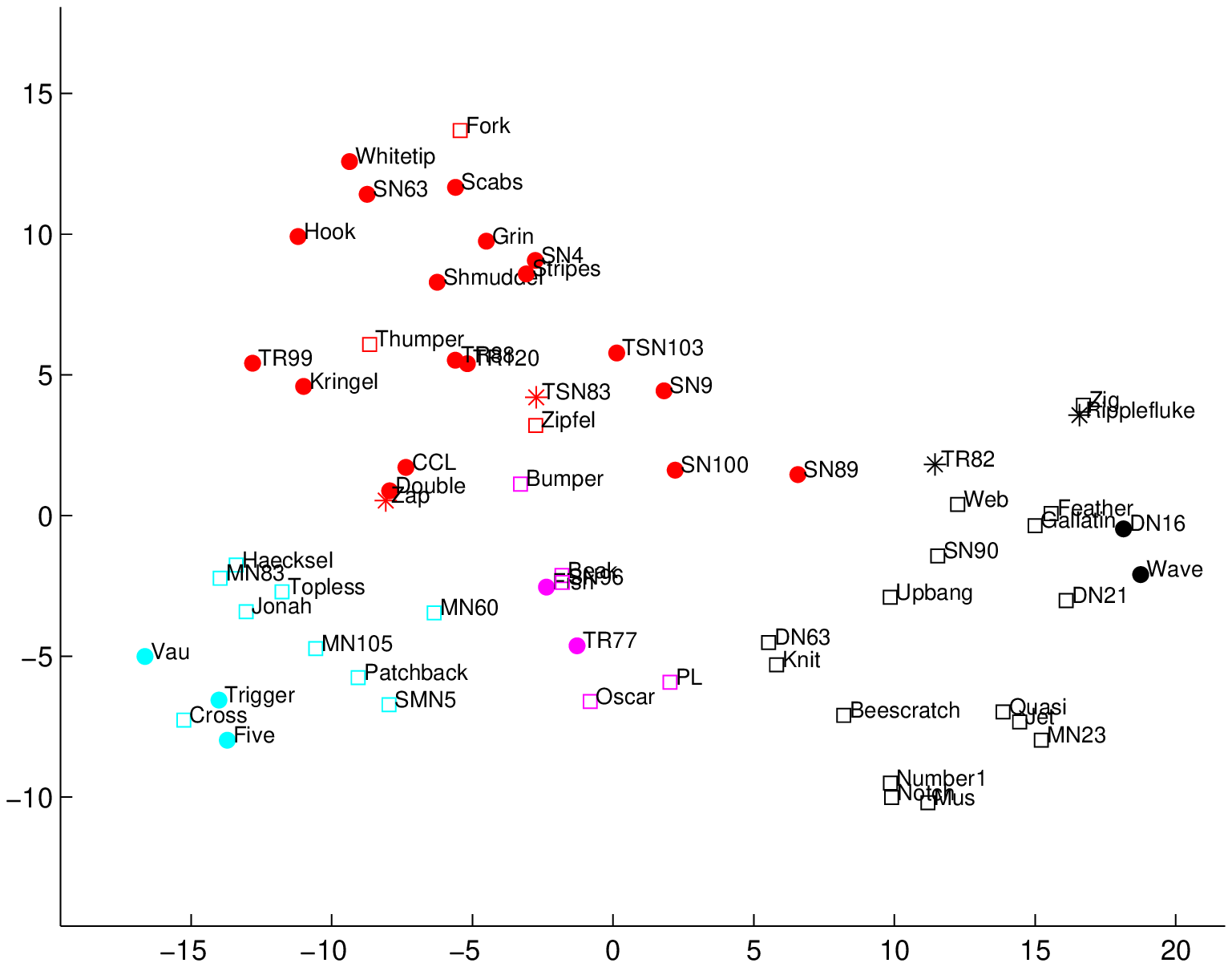}  \\
	 (a)& (b)
	\end{tabular}
	\begin{tabular}{c}
      \epsfxsize 12cm \epsfbox{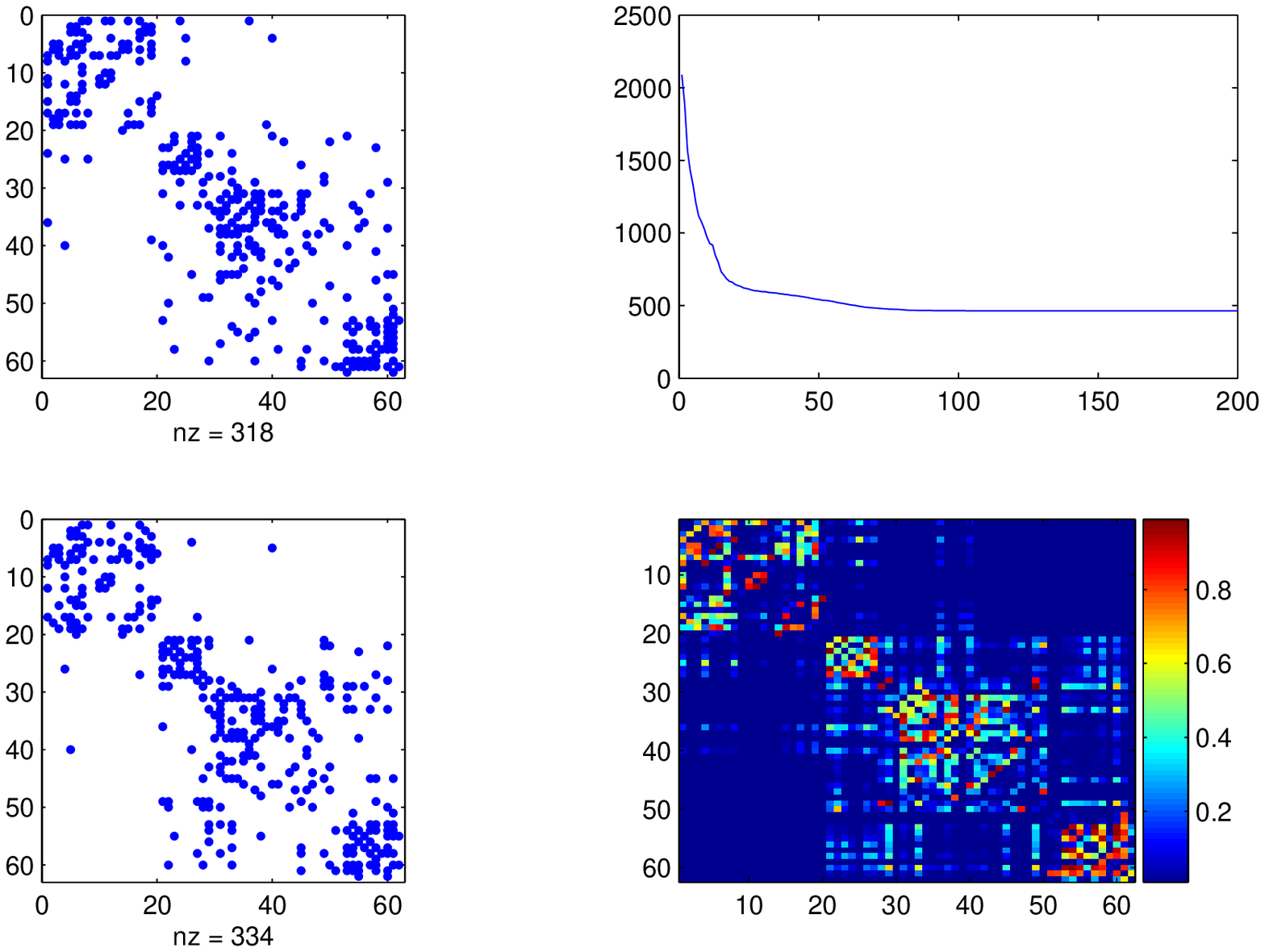}\\
	(c)
	\end{tabular}
	\end{center}
	\caption{(a) Dolphins social network. (b) Distribution of attributes.
			(c) Fitted probabilistic model of Karate clube network. Upper left: adjacency matrix of dolphins network. Upper right: decay of fitting error when training. Lower left: adjacency matrix of a surrogate network generated by fitted model. Lower right: the fitted probabilistic model.}
	\label{fig:fig6}
\end{figure}

\begin{figure}[ttbb]
   	\begin{center}
	\begin{tabular}{c}
	\epsfxsize 10cm \epsfbox{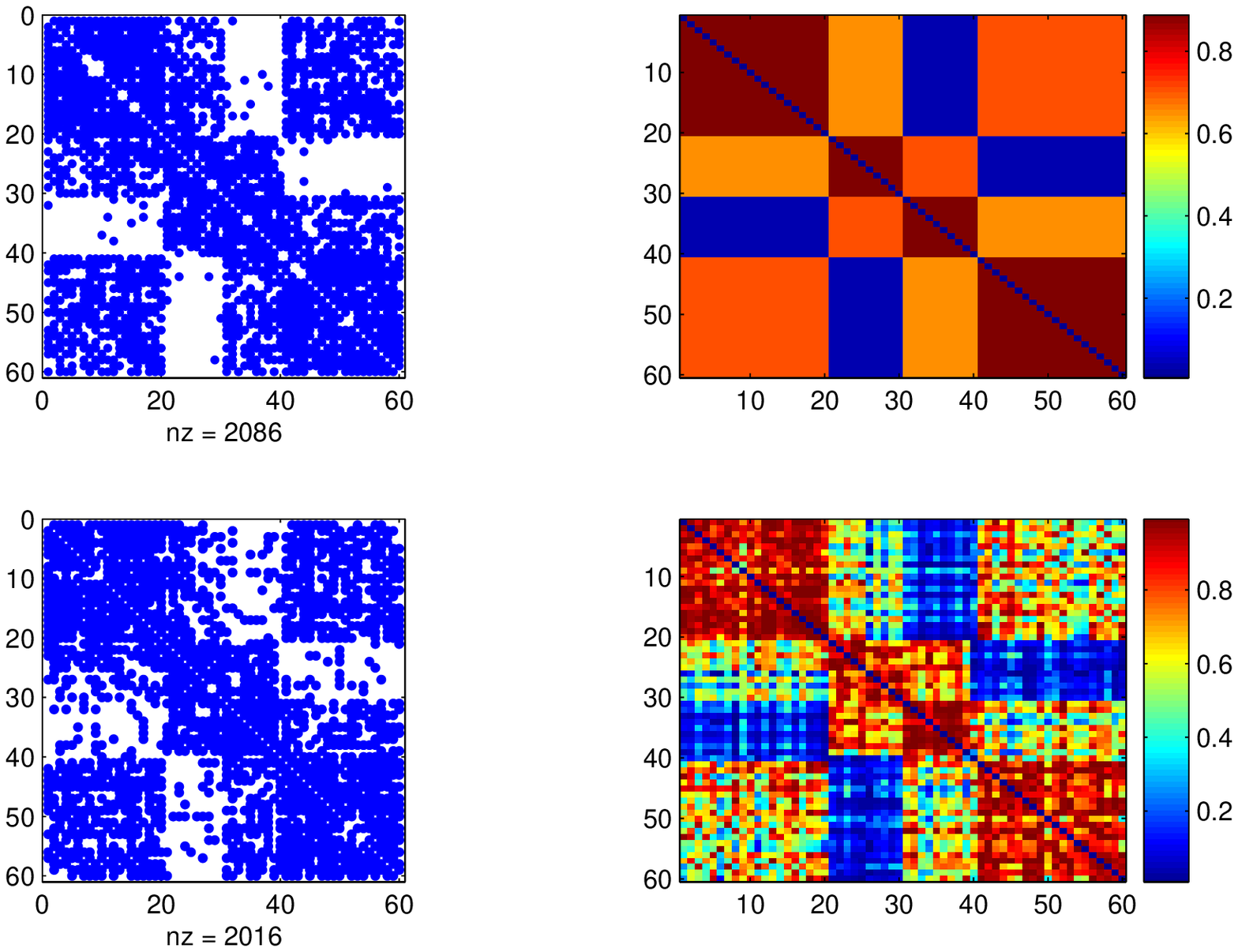} \\
	(a)\\
	\epsfxsize 10cm \epsfbox{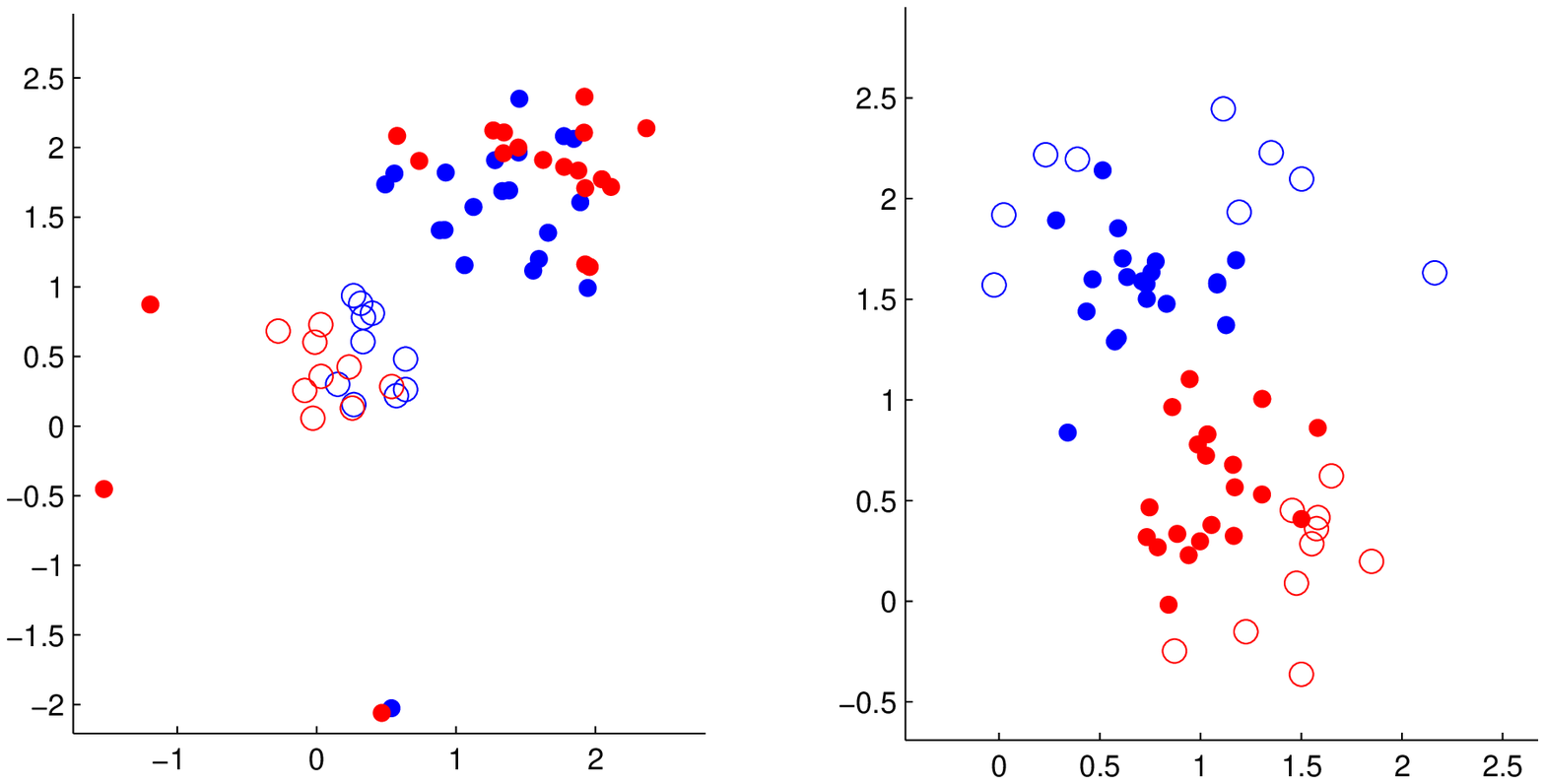}\\
	(b)\\
	\end{tabular}
	\end{center}
	\label{fig:fig7}
	\caption{ (a) Multi-layered clustering netowrk. Upper left: adjancency matrix of observed 		network. Upper right: true unerlying generative model. Lower left: adjancency matrix 		of surrogate network. Lower right: fitted probabilistic model. 
		(b) Distribution of attributes for $2$-layered clustered network. Left: $w^{(1)}(1)$ vs. $w^{(1)}(2)$. Right: $w^{(2)}(1)$ vs. $w^{(2)}(2)$. In both figures, different colors and different symbols are used to indicate which groups in different layer the nodes belong to. E.g., nodes in blue color (no.$31$ to $60$) are in one cluster and nodes in red color (no.$1$ to $30$) are in another cluster at first layer. Similarly, the dotted nodes (no.$10$ to $20$ and no.$41$ to $60$) and circled nodes (no.$21$ to $40$) are in different clusters at second layer.}
\end{figure}

American college football teams network has many clusters involved. The fitted model nevertheless successfully capture this complicated structure. The amazing representation of the network in atribute space (shown in figure 5(a)) not only manifests clustering structures, but also suggests that several wandering points may not be clearly classfied. It turns out that these nodes belongs to specific group (IA Independents) and can not be classifed as one cluster consistently. Furthermore, the particular node in the group of Conference USA which is classifed as member of the group of Western Athletic is not due to the flaw of the method but caused by the network construction\cite{Girvan2002}. The attributes representation also suggests some groups may be further divided into smaller subgroups such as the Mid American group. This observation is well supported by the connection patterns of nodes in this group. These useful information can not be easily obtained by conventional network clustering algorithms, and show the advantages of the proposed modeling approach.

The dolphins social network is another widely studied example. It can be divided into two big clusters. One of the cluster may be further divided into three small group as studied in \cite{Lusseau2004}. The analysis results for dolphins social network are shown in figure 6. The distribution of the attributes correctly reflects $2$ large clusters (as shown in figure 6(b)), but also indicates that there may exists $2$ or $3$ smaller clusters in the left cluster. 
The interesting observation is that the attributes of the nodes corresponding to three small clusters suggested in \cite{Lusseau2004} are indeed well grouped. Also, the surrogate network generated by the fitted model (as shown in figure 6(c)) shows striking similarity with the observed network.

Decaying behaviors of fitting errors when the optimization procedure iterated unveil some common feature, which can be seen from the subplots (upper right ones) in all examples studied as shown in figure 4(c), figure 5(c) and figure 6(c). It can be divided into two segments, a rapid decrease phase and a gradual change phase. After closely monitoring the movement of the points in attribute space step by step, we find that the points are always firstly arranged according to globe structure of the network, and then fine adjusted within each cluster to generate better configuration. This may partially explained why the suboptimal solutions due to gradient based algorithm also give out good configuration in general. 

Although modular structures are most commonly observed and studied, there are many different other structures. In our study, we choose Gaussian function to relate local attributes to connection probability, which is particular useful to capture clustering structure. The proposed modeling approach nevertheless is flexible enough to deal with other structures providing they can be well defined. For example, for a bipartite network, the function $1-f$, where $f$ is similar to what used in the paper, may be a better choise.

The proposed modeling approach can be easily extended to deal with more complicated grouping structure than simple modular structure. Let us consider a simple exention of overlapping two clustering structure. For example, consider a group of students who make friends based on different factors such as personality or avocation. Now even if each friendship network based on any single factor were well clustered, the observed overall network may have structure quite different from the simple modular network. We call this kind of network multi-layered modular network. One such example of $2$-layered network is shown in figure 7(a). This particular network is generated by the following way. Suppose the nodes are numbered from $1$ to $60$. We first generate two modular networks separately. The first modular network has two clusters, one containing node $1$ to node $30$ and the other containing node $31$ to node $60$. The connection probabilities are $0.7$ for nodes in same cluster and $0.02$ in different clusters. The second modular network also has two clusters, one containing node $21$ to node $40$, and the other containing nodes $1$ to $20$ and nodes $41$ to $60$. The connection probabilities are slightly different, which are $0.6$ for nodes in same cluster and $0.01$ for nodes in different clusters. The final network is generated by stacking two modular networks together, and removing repeated links. 
To explore such multi-layered structure by the proposed modeling approach, the most straight way is to extend the dimension of the attributes vector. Let $w=[w^{(1) T},w^{(2) T}]^T$ the new extended attributes. For each layer, the connection probability can be written as $p^{(1)}_{ij} \sim f(\parallel w^{(1)}_i -w^{(1)}_j \parallel)$ and $p^{(2)}_{ij} \sim f(\parallel w^{(2)}_i -w^{(2)}_j \parallel)$.
The connection probability of the observed network will be $p_{ij}=p^{(1)}_{ij}+p^{(2)}_{ij}-p^{(1)}_{ij}p^{(2)}_{ij}$. Following the same procedure described above, we get the attributes representation in extended attribute space. The distribution of the attributes are shown in figure 7(b). Interestingly, the hidden clustering structure are unveiled in different subspace of the extended attribute space. 

In summary, we proposed a probabilistic modeling approach to analyze networks. Under this framework, the observed network can be regarded as a measurement of certain probabilistic system, where the connection probability of any pair of nodes depends on the properly rescaled distance between the introduced local attributes of the corresponding nodes. It is remarkable that the configuration of the optimally estimated attributes well represents the intrinsic structure of the observed network, thus provides an very informative way to visualize networks in low-dimensional space. It can be more effective to make further network structure analysis based on the attribute vectors instead of observed network directly. The modeling approach can be easily extended to deal with more complicated structures such as multi-layered clustered network.

\section{Acknowledgment}

This work is supported by Temasek Laboratories at National University of Singapore through the DSTA Project POD0613356.

\end{document}